\documentstyle[12pt]{article}

\newcommand{\be}{\begin{equation}}
\newcommand{\ee}{\end{equation}}
\newcommand{\bea}{\begin{eqnarray}}
\newcommand{\eea}{\end{eqnarray}}
\newcommand{\ba}{\begin{array}}
\newcommand{\ea}{\end{array}}
\newcommand{\non}{\nonumber}
\newcommand{\PP}[1]{\; P^{(#1)}}
\newcommand{\pp}[2]{\; p^{(#1,#2)}}
\newcommand{\eee}[2]{\; e^{(#1,#2)}}
\newcommand{\bb}[2]{\; b^{(#1,#2)}}
\newcommand{\bbp}[2]{\; {b^{+}}^{(#1,#2)}}
\newcommand{\bbm}[2]{\; {b^{-}}^{(#1,#2)}}
\newcommand{\bbpm}[2]{\; {b^{\pm}}^{(#1,#2)}}
\newcommand{\id}{\mbox{$I$}}

\newcommand{\BMA}{braid--monoid algebra}
\newcommand{\BWM}{Birman--Wenzl--Murakami}
\newcommand{\BWMA}{Birman--Wenzl--Murakami algebra}

\newcommand{\YBE}{Yang--Baxter equation}
\newcommand{\YBO}{Yang--Baxter operator}

\newcommand{\zr}[1]{\mbox{\hspace*{#1em}}}
\newcommand{\CC}{\mbox{{\sf C}\zr{-0.35}\rule{0.04em}{1.6ex}\zr{0.35}}}
\newcommand{\dys}{\displaystyle}
\newcommand{\dpo}{\mbox{$(d\! +\! 1)$}}
\newcounter{eqs}
\newcommand{\scestart}{\setcounter{eqs}{\theequation}
   \addtocounter{eqs}{1}
   \setcounter{equation}{0}
   \renewcommand{\theequation}{\arabic{eqs}\alph{equation}}}
\newcommand{\sceend}{\setcounter{equation}{\theeqs}
   \renewcommand{\theequation}{\arabic{equation}}}
\newcommand{\scestep}{\addtocounter{eqs}{1}
                      \setcounter{equation}{1}}

\begin{document}
\begin{center}
{\LARGE\bf Trigonometric $R$ Matrices related to \\[4mm]
           ``Dilute'' \BWM\ Algebra} \\[8mm]
{\large\sc Uwe Grimm} \\[6mm]
{\footnotesize Instituut voor Theoretische Fysica,
               Universiteit van Amsterdam, \\
               Valckenierstraat 65,
               1018 XE Amsterdam,
               The Netherlands} \\[1mm]
{\footnotesize Email: grimm@phys.uva.nl}\\[6mm]
February 1994 \\[6mm]
{University of Amsterdam preprint ITFA--94--02} \\[6mm]
\end{center}
\begin{quote}
{\small\bf Abstract.}
{\small\sf
 Explicit expressions for three series of $R$ matrices which
 are related to a ``dilute'' generalisation of the
 \BWMA\ are presented. Of those, one series is equivalent to
 the quantum $R$ matrices of the $D^{(2)}_{n+1}$ generalised Toda
 systems whereas the remaining two series appear to be new.
}
\end{quote}\vspace*{8mm}

A ``dilute'' generalisation of the
\BWM\ (BWM) algebra \cite{BirWen,Mura} has recently been
introduced \cite{Gri94a,GriPea93}. It appears \cite{Gri94a}
as a generalised \BMA\ \cite{WadDegAku}
related to certain exactly solvable lattice models of
two-dimensional statistical mechanics.
Alternatively, it can be regarded as a particularly simple
case of a two-colour \BMA\ \cite{GriPea93} where one colour
is represented trivially
(in the sense that the corresponding representation
of the subalgebra generated by the elements of this colour
is one-dimensional).
In Ref.~\cite{Gri94a}, it was shown that representations
of this algebra can be ``Baxterised'' \cite{Jones},
i.e., one can find a general expression for
a local \YBO\ $X_{\, j}(u)$ \cite{WadDegAku}
($u$ denoting the spectral parameter)
in terms of the generators of the dilute BWM algebra for
which the Yang--Baxter relations
\be
\ba{l}
X_{\, j}(u)\: X_{\, j+1}(u+v)\: X_{\, j}(v) \;\; = \;\;
X_{\, j+1}(v)\: X_{\, j}(u+v)\: X_{\, j+1}(u)  \\[1mm]
X_{\, j}(u)\: X_{\, k}(v) \;\; = \;\; X_{\, k}(v)\: X_{\, j}(u)
\hspace*{15mm} \mbox{for $|j-k|>1$} \ea
\label{YBA}
\ee
follow algebraically. This implies that every suitable
representation of the dilute BWM algebra defines a
solvable lattice model. As an example, one series of $R$
matrices of this kind has been given in Ref.~\cite{Gri94a}
which were shown to be equivalent to the $D^{(2)}_{n+1}$
vertex models \cite{Bazh85,Jimbo86}.

In this letter, we present the explicit form of
three such series of $R$ matrices (where the series
mentioned above is included for completeness).
In this case, the \YBO\ $X_{\, j}(u)$ acts on a tensor space
\mbox{$V\otimes V\otimes\ldots\otimes V$}
(where $V\cong{\CC}^{d+1}$ with some integer $d$) as
\be
X_{\, j}(u) \;\; = \;\; \id\otimes\id\otimes\ldots\otimes\id\otimes
\check{R}(u)\otimes\id\otimes\ldots\otimes\id\otimes\id
\label{X} \ee
where $\check{R}(u)=PR(u)$ ($P$ is the permutation map
on $V\otimes V$, i.e., $P$: $v\otimes w\mapsto w\otimes v$)
acts on the $j$ and \mbox{$j\! +\! 1$} factors in the tensor product
and $R(u)$ is the corresponding $R$ matrix.

The representations of the
dilute BWM algebra considered in what follows
can be regarded as the ``dilutisation'' of
well-known representations of the BWM algebra itself.
These are the representations which describe the
$B^{(1)}_{n}$, $C^{(1)}_{n}$ and $D^{(1)}_{n}$ vertex models
\cite{Bazh85,Jimbo86,DegWadAku88b,CheGeLiuXue92}
(and the $A^{(2)}_{n}$ models as well, see e.g.~\cite{Gri94a}),
we will refer to the corresponding series of (dilute) models
as the (B), (C), and (D) series for short.
The corresponding representations of the dilute algebra
are obtained by adding a single state
(which carries the second (``trivial'') colour)
to the local space
$\tilde{V}\cong{\CC}^{d}$ (with $d=2n+1$ for the (B) and
$d=2n$ for the (C) and (D) series, respectively) yielding
$V=\tilde{V}\oplus{\CC}\cong{\CC}^{d+1}$.

For all three series,
the \mbox{${\dpo}^2\times {\dpo}^2$} matrix
$\check{R}(u)$ is given by the following general
expression \cite{Gri94a}
\bea
\check{R}(u) & = &
 \; \pp{1}{1} \non\\* & &
\mbox{} + \; \zeta^{-1}\,\eta^{-1}\,(z-z^{-1})\,
  \left(\tau^{-1}z\bbp{1}{1}\; -\tau z^{-1}\bbm{1}{1}\;\right) \non\\* & &
\mbox{} + \; \eta^{-1}\, (\tau z^{-1}-\tau^{-1}z)\,
  \left(\pp{1}{2}\; +\pp{2}{1}\;\right) \non\\* & &
\mbox{} - \; \kappa_{1}\,
  \zeta^{-1}\,\eta^{-1}\, (z-z^{-1})\, (\tau z^{-1}-\tau^{-1}z)\,
  \left(\bb{1}{2}\; +\bb{2}{1}\;\right) \non\\* & &
\mbox{} + \; \kappa_{2}\,\eta^{-1}\, (z-z^{-1})\,
  \left(\eee{1}{2}\; +\eee{2}{1}\;\right) \non\\* & &
\mbox{} + \; \left(1\: -\: \zeta^{-1}\,\eta^{-1}\, (z-z^{-1})\,
           (\tau z^{-1}-\tau^{-1}z)\right)\, \pp{2}{2}
\label{R}
\eea
where \mbox{$z=\exp(iu)$},
\mbox{$\zeta=(\sigma-\sigma^{-1})$},
\mbox{$\eta=(\tau-\tau^{-1})$}
and where one can choose arbitrary signs
\mbox{$\kappa_{1}^2=\kappa_{2}^2=1$} (cf.\ Ref.~\cite{Gri94a}).
Here, the
relation between $\sigma$  and $\tau$
is given by
\mbox{$\tau^2=\sigma^{2n},-\sigma^{2n+1},\sigma^{2n-1}$}
for the (B), (C), and (D) series, respectively.

Following closely the notation of Ref.~\cite{Jimbo86}
(with $\sigma=k$ and $\tau^2=\xi$), we find the following
explicit expressions for the matrices
\mbox{$\pp{a}{b}=\PP{a}\otimes\PP{b}$},
$\bbpm{a}{b}$, and $\eee{a}{b}$
\mbox{($a,b\in\{ 1,2\}$)}\scestart
\bea
\PP{1} & = & \sum_{\alpha} \: E_{\alpha,\alpha} \label{repp1} \\
\PP{2} & = & E_{d+1,d+1} \label{repp2} \\[4mm] \scestep
\bbp{1}{1} & = &  \;
\sum_{\alpha}\: \sigma^{-1}\: \left(1\: +\: (\sigma-1)\,
\delta_{\alpha,\alpha^{\prime}}\right)\:
E_{\alpha,\alpha}\otimes E_{\alpha,\alpha} \non\\* & &
\mbox{} +\; \sum_{\alpha\neq\beta}\: \left(1\: +\: (\sigma-1)\,
\delta_{\alpha,\beta^{\prime}}\right) \:
E_{\alpha,\beta}\otimes E_{\beta,\alpha} \non\\* & &
\mbox{} -\; (\sigma-\sigma^{-1}) \:\sum_{\alpha<\beta}\:
E_{\alpha,\alpha}\otimes E_{\beta,\beta} \non\\* & &
\mbox{} +\; (\sigma-\sigma^{-1}) \:\sum_{\alpha>\beta}\:
\varepsilon_{\alpha}\:\varepsilon_{\beta}\:
\sigma^{\bar{\alpha}-\bar{\beta}}\:
E_{\alpha^{\prime},\beta}\otimes E_{\alpha,\beta^{\prime}}
\label{repb1} \\
\bbm{1}{1} & = & \;
\sum_{\alpha}\: \sigma\: \left(1\: +\: (\sigma^{-1}-1)\,
\delta_{\alpha,\alpha^{\prime}}\right)\:
E_{\alpha,\alpha}\otimes E_{\alpha,\alpha} \non\\* & &
\mbox{} +\; \sum_{\alpha\neq\beta}\: \left(1\: +\: (\sigma^{-1}-1)\,
\delta_{\alpha,\beta^{\prime}}\right) \:
E_{\alpha,\beta}\otimes E_{\beta,\alpha} \non\\* & &
\mbox{} +\; (\sigma-\sigma^{-1}) \:\sum_{\alpha>\beta}\:
E_{\alpha,\alpha}\otimes E_{\beta,\beta} \non\\* & &
\mbox{} -\; (\sigma-\sigma^{-1}) \:\sum_{\alpha<\beta}\:
\varepsilon_{\alpha}\:\varepsilon_{\beta}\:
\sigma^{\bar{\alpha}-\bar{\beta}}\:
E_{\alpha^{\prime},\beta}\otimes E_{\alpha,\beta^{\prime}}
\label{repb2} \\
b^{(1,2)} & = & \sum_{\alpha}\:
E_{d+1,\alpha}\otimes E_{\alpha,d+1} \label{repb3} \\
b^{(2,1)} & = & \sum_{\alpha}\:
E_{\alpha,d+1}\otimes E_{d+1,\alpha} \label{repb4} \\[4mm] \scestep
e^{(1,2)} & = & -\: \sum_{\alpha}\:
\varepsilon_{\alpha}\:\sigma^{(d+1)/2-\bar{\alpha}}\:
E_{d+1,\alpha}\otimes E_{d+1,\alpha^{\prime}} \label{repe1} \\
e^{(2,1)} & = & -\: \sum_{\alpha}\:
\varepsilon_{\alpha}\:\sigma^{\bar{\alpha}-(d+1)/2}\:
E_{\alpha^{\prime},d+1}\otimes E_{\alpha,d+1}
\label{repe2}
\eea
where\sceend in all expressions the summation variables
are restricted to values \mbox{$1\leq\alpha,\beta\leq d$}.
Here, $E_{k,\ell}$ denote \mbox{$\dpo\times\dpo$}
matrices with elements
\mbox{${(E_{k,\ell})}_{i,j}=\delta_{i,k}\delta_{j,\ell}$}.
Furthermore, ``charge conjugated'' states
are defined by \mbox{$\alpha^{\prime}=d+1-\alpha$}
\mbox{($1\leq\alpha\leq d$)} and
\mbox{${\dpo}^{\prime}=\dpo$}.
We use \mbox{$\varepsilon_{\alpha}=1$} for the (B) and (D) series
whereas
\be
\varepsilon_{\alpha} \;\; = \;\; \left\{
\ba{r@{\hspace*{10mm}}l}
1  & \alpha<\alpha^{\prime} \\
-1 & \alpha>\alpha^{\prime}
\ea \right.
\label{eps}
\ee
for the (C) series.
Finally,
\be
\bar{\alpha} \;\; = \;\; \left\{
\ba{l@{\hspace*{5mm}}l}
\alpha\;\pm\;\frac{1}{2} & \alpha<\alpha^{\prime} \\
\alpha                   & \alpha=\alpha^{\prime} \\
\alpha\;\mp\;\frac{1}{2} & \alpha>\alpha^{\prime}
\ea \right.
\label{bar}
\ee
where the upper sign applies for series (B) and (D)
and the lower for the (C) series, respectively.

It is straightforward to show that the above expressions
(\ref{repp1})--(\ref{repe2}) indeed define a matrix representation
of the dilute BWM algebra in the sense of Ref.~\cite{Gri94a}.
It follows from the results of Ref.~\cite{Gri94a} that the
$\check{R}$ matrix (\ref{R}) fulfills the quantum \YBE\ (\ref{YBA})
\bea
\lefteqn{
\left(\check{R}(u)\otimes\id\right) \:
\left(\id\otimes\check{R}(u+v)\right) \:
\left(\check{R}(v)\otimes\id\right) } \non \\*
& \hspace*{10mm} & = \;\;
\left(\id\otimes\check{R}(v)\right) \:
\left(\check{R}(u+v)\otimes\id\right) \:
\left(\id\otimes\check{R}(u)\right) \;\; .
\eea
In addition, it satisfies the following relations\scestart
\bea
\makebox[80mm][l]{$\dys\check{R}(0) \;\; = \;\; \id$}
 & \hspace*{10mm} &
\mbox{\small (initial condition)} \label{prop1} \\*
\makebox[80mm][l]{$\dys\check{R}(u)\:\check{R}(-u) \;\; = \;\;
\varrho(u)\:\varrho(-u) \;\id$} & \hspace*{10mm} &
\mbox{\small (inversion relation)} \label{prop2} \\*
\makebox[80mm][l]{$\dys\check{R}(u) \;\; = \;\;
\mbox{}^{t}\check{R}(u)$}  & \hspace*{10mm} &
\mbox{\small (reflection symmetry)\hspace*{10mm}} \label{prop3} \\*
\makebox[80mm][l]{$\dys
\check{R}^{\, k\, p\,}_{\, \ell\, q\,}(u)\;\; = \;\; 0$
\hspace*{5mm} unless $\tilde{k}+\tilde{\ell}=\tilde{p}+\tilde{q}$}
& \hspace*{10mm} & \mbox{\small (charge conservation)} \label{prop4} \\*
\makebox[80mm][l]{$\dys
R(u) \;\; = \;\; (S\otimes S)\;\: {}^{t}R(u)\;\: (S\otimes S)$}
& \hspace*{10mm} & \mbox{\small (CT invariance)} \label{prop5} \\*
\makebox[80mm][l]{$\dys
R(u) \;\; = \;\;
(C\otimes\id)\;\: {}^{t_{1}}(P\, R(\lambda-u)\, P)\;\:
{(C\otimes\id)}^{-1}$}
& \hspace*{10mm} &  \mbox{\small (crossing symmetry)}  \label{prop6}
\eea
where\sceend relations (\ref{prop3})
(P invariance) and (\ref{prop5}) (CT invariance)
do {\em not} hold for the (C) series, but the
invariance under the combined operation (CPT invariance)
persists. Here, left superscripts
$t_{1}$ and $t$  denote transposition in the first
space and in both spaces, respectively.
The function $\varrho(u)$ which
enters in the inversion relation (\ref{prop2}) has the form
\be
\varrho(u) \;\; = \;\;  \zeta^{-1}\,\eta^{-1}\,
(\sigma z^{-1}-\sigma^{-1}z)\, (\tau z^{-1}-\tau^{-1} z) \;\; .
\label{rho}
\ee
The matrix elements of $\check{R}(u)$ are defined by
\be
\check{R}(u) \;\; = \;\; \sum_{k,\ell,p,q=1}^{d+1}\:
\check{R}^{\, k\, p\, }_{\,\ell\, q\, }(u) \:
E_{q,k}\otimes E_{p,\ell}
\label{RR}
\ee
and $\tilde{k}$ is given by
\be
\tilde{k} \;\; = \;\; \frac{k-k^{\prime}}{2} \;\; .
\label{til}
\ee
Note that the state \mbox{$\:d\! +\! 1\:$}
is charge conjugated to itself,
i.e., \mbox{${\dpo}^{\prime}=\dpo$} and hence
\mbox{$\widetilde{d\! +\! 1}=0$}.
Finally, the crossing parameter $\lambda$ is determined by
\mbox{$\tau=\exp(i\lambda)$} and the matrices $S$ and $C$ in
Eqs.~(\ref{prop5}) and (\ref{prop6}) have elements
\be
C_{i,j} \;\; = \;\;
r(i)\; S_{i,j} \;\; = \;\;
r(i)\;\delta_{i,j^{\prime}}
\ee
with crossing multipliers
\be
r(\alpha) \;\; = \;\; \varepsilon_{\alpha}\:
\sigma^{\bar{\alpha}-(d+1)/2}\;\; ,  \hspace*{10mm}
r(d+1)\;\; = \;\; -\kappa_{2} \;\; .
\ee

The crossing symmetry (\ref{prop6}) of the $\check{R}$ matrix
(\ref{R}) can easily be
verified by looking at the properties of the individual
parts (\ref{repp1})--(\ref{repe2}) under the
``crossing transformation''
\be
{\cal O}\longmapsto {\cal O}^{cr} =
P\:(C\otimes\id)\;\:{}^{t_{1}}({\cal O}\, P)\;\: {(C\otimes\id)}^{-1}
\label{crosstrafo}
\ee
where ${\cal O}$ denotes any of the matrices of
Eqs.~(\ref{repp1})--(\ref{repe2}).
For \mbox{$\kappa_{2}=1$} (see Ref.~\cite{Gri94a}),
one observes\scestart
\bea
\pp{1}{1}\;\; & \longmapsto & \eee{1}{1}\;\;
\;\;\longmapsto\;\; \pp{1}{1} \\*
\bbp{1}{1} & \longmapsto & \bbm{1}{1}
\;\;\longmapsto\;\; \bbp{1}{1} \\*
\pp{1}{2}\;\; & \longmapsto & \eee{1}{2}\;\;
\;\;\longmapsto\;\; \pp{2}{1}\;\;
\;\;\longmapsto\;\; \eee{2}{1}\;\;
\;\;\longmapsto\;\; \pp{1}{2} \\*
\bb{1}{2}\;\; & \longmapsto & \bb{2}{1}\;\;
\;\;\longmapsto\;\; \bb{1}{2} \\*
\pp{2}{2}\;\; & \longmapsto & \pp{2}{2}
\eea
where\sceend
\mbox{$\eee{1}{1}=\id\otimes\id+\xi^{-1}(\bbp{1}{1}-\bbm{1}{1})$}.
This is exactly what one expects from the
diagrammatic interpretation of the corresponding
generators of the ``dilute'' BWM algebra (in which the
crossing transformation (\ref{crosstrafo}) corresponds to
a clockwise rotation of the two-string diagrams by 90 degrees),
compare the discussion in Ref.~\cite{GriPea93}.

The $R$ matrices defined in this letter actually possess
a charge conservation that is somewhat stricter
than shown in Eq.~(\ref{prop4}). In fact, the charges of
both ``colours'' are conserved separately which
implies that only vertices with an even number of states
\mbox{$\: d\! +\! 1\: $} occur (i.e., the number of
states \mbox{$\: d\! +\! 1\: $} is conserved modulo two
as this state is conjugated to itself).

As shown explicitly in Ref.~\cite{Gri94a},
the (B) series is equivalent to the $D^{(2)}_{n+1}$
vertex models, differing from the $R$ matrix of Ref.~\cite{Jimbo86}
by a local orthogonal transformation only.
Essentially, the basis of Ref.~\cite{Jimbo86} uses the
symmetric and anti-symmetric linear combinations
(states \mbox{$\: n\! +\! 1\: $} and
\mbox{$\: n\! +\! 2\:$} in Ref.~\cite{Jimbo86}, respectively)
of the two ``neutral'' states
\mbox{$\: (d\! +\! 1)/2=n\! +\! 1\:$} and
\mbox{$\: d\! +\! 1=2n\! +\! 2\:$}
in the present parametrisation.
For the other two series, it remains an open question
if similar relations to known $R$ matrices exist.

This work was supported by a Fellowship of the
European Community (Human Capital and Mobility Programme).
The author thanks P.\ A.\ Pearce and
S.\ O.\ Warnaar for valuable comments.

\end{document}